\documentclass{sig-alt-release2}

\title{Towards an automated query modification assistant}

 \numberofauthors{2} 
 \author{
 \alignauthor
 Vera Hollink\\
        \affaddr{Centre for Mathematics and Computer Science}\\
        \affaddr{Interactive Information Access Group}\\
        \affaddr{Amsterdam, The Netherlands}\\
        \email{V.Hollink@cwi.nl}
 \alignauthor
 Arjen de Vries\\
        \affaddr{Centre for Mathematics and Computer Science}\\
        \affaddr{Interactive Information Access Group}\\
        \affaddr{Amsterdam, The Netherlands}\\
        \email{arjen@acm.org}
 }
 
 \makeatletter
\let\@copyrightspace\relax
\makeatother

\begin{document}
\maketitle

\category{H3.3}{Information Search and Retrieval}{Query formulation}
\category{H3.3}{Information Search and Retrieval}{Search process}

\terms{Experimentation, Algorithms}


\begin{abstract}
Users who need several queries before finding what they need can benefit from an automatic search assistant that provides feedback on their query modification strategies. We present a method to learn from a search log which types of query modifications have and have not been effective in the past. The method analyses query modifications along two dimensions: a traditional term-based dimension and a semantic dimension, for which queries are enriches with linked data entities. Applying the method to the search logs of two search engines, we  identify six opportunities for a query modification assistant to improve search: modification strategies that are commonly used, but that often do not lead to satisfactory results.
\end{abstract}

\section{Introduction}
Users of search engines often enter a number of queries in succession before they find everything they need or before they are convinced that the collection in which they search does not contain answers to their information needs. Query suggestions can help users in the formulation of their queries  (e.g. \cite{Cao08context, Kelly09comparison}).
In addition to query suggestions, users can potentially be helped by higher level feedback on their search strategy.  Such feedback can warn users earlier on when the current line of search is not going to be effective and help them to formulate a more effective strategy. For instance, suppose a users is looking for pictures of impressionist paintings. She successively enters the queries \texttt{Monet} and \texttt{impressionism}, but none of the search results are of her liking. Query suggestions may include the names of contemporaries of Monet and the titles of paintings by Monet. This can be useful suggestions, but do not tell the user why her own modified query (\texttt{impressionism}) was unsuccessful. Feedback on her modification strategy can provide the answer:
\begin{quotation} 
`When the name of an artist does not yield any relevant results, then entering the name of the style used by the artist usually does not yield relevant results either. Instead you could try searching on the names of other artists using this style or titles of paintings in this style.'  
 \end{quotation}
This type of feedback offers two advantages over query suggestions alone: 
\begin{enumerate}
 \item It provides directions for formulating queries even when the offered query suggestions do not appeal to the user.
 \item The suggested search strategies can be reused for comparable information needs (e.g. finding cubist paintings), preventing the user from making the same `mistake' twice.
\end{enumerate}

In this paper we explore the potential of an automatic search assistant that provides feedback on the ways users modify their queries. We present a methodology to use search interactions of previous users collected in search logs to determine which types of query modification have and have not been effective in the past. These observations can form the basis for the assistant's feedback. In addition, we identify opportunities where the assistant may improve the search: query modification strategies that are commonly used, but do not usually lead to results that users find relevant.

We analyze query modifications in two ways: by a traditional term-based approach (e.g. \cite{Boldi09dango,Costa08hyponymy,He02combining,Huang09analyzing,Jansen09patterns,Jorgensen05image,Ozmutlu09markovian,Whittle07data}) and by a semantic approach that we developed for this purpose \cite{Hollink11semantic}. Both approaches are applied to query logs of the search engines of two image repositories. 
 
The remainder of this paper is organized as followes. In Section~\ref{relatedWork} we discuss related studies on query modifications. The analysis method is presented in Section~\ref{method}. Section~\ref{dataSets} introduces the data sets that are used in the experiments in Section~\ref{results}. The last section contains conclusions and discussed our results.

\section{Related work} \label{relatedWork}
\begin{figure*}[t]
\centering
\includegraphics[width=0.9\textwidth]{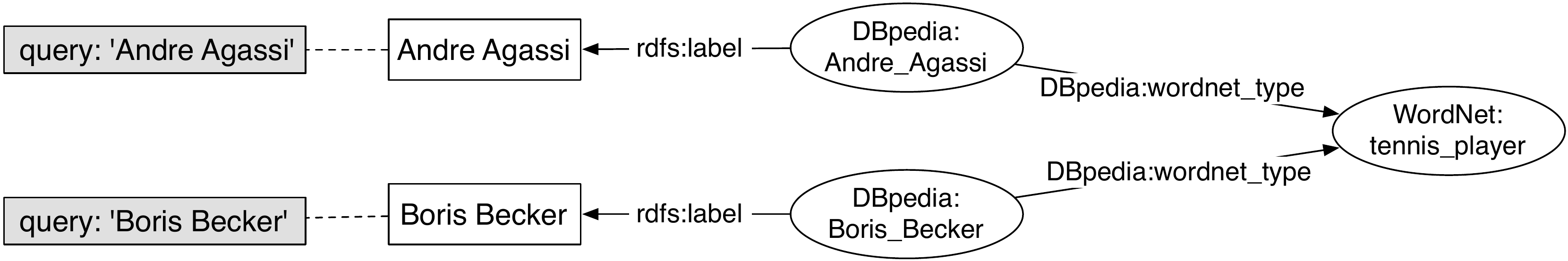}
\caption{Example application of our method for finding semantic relations between queries: a relation between queries \texttt{Andre Agassi} and \texttt{Boris Becker} is that they both match DBpedia entities that are of WordNet type \texttt{tennis\_player}.}
\label{pathSearchExampleFig}
\end{figure*}

Research on query modifications studies pairs of queries that are successively submitted in a search session (e.g. \cite{Boldi09dango,Costa08hyponymy,He02combining,Huang09analyzing,Jansen09patterns,Jones03query,Jorgensen05image,Ozmutlu09markovian,Rieh06analysis,Whittle07data}.) Successive query pairs are classified into a number of modification classes and the use of these classes is analyzed. 

Query modifications are classified either manually \cite{Jones03query,Jorgensen05image,Rieh06analysis} or automatically \cite{Boldi09dango,Costa08hyponymy,He02combining,Huang09analyzing,Jansen09patterns,Ozmutlu09markovian,Whittle07data}. Studies that employ automatic methods usually classify query modifications solely on the basis of terms in the queries. These studies examine whether terms have been added, eliminated or substituted compared to the user's previous query. When terms are added, the modification is classified as \emph{specification} (e.g., from query \texttt{Beckham} to query \texttt{Beckham Milan}), when terms are eliminated it is classified as \emph{generalization} (from \texttt{Beckham Milan} to \texttt{Beckham}), and when terms are substituted it is classified as \emph{reformulation} (from \texttt{Beckham Milan} to \texttt{Beckham Madrid}). Finally, \emph{lexical variations} include, for instance, modifications from singular to plural forms or vice versa. In some of the manual studies not only term overlap, but also the meaning of the queries is taken into account \cite{Rieh06analysis,Jorgensen05image,Jones03query}. In these studies the same modification classes are used, for instance, a modification from \texttt{dog} to \texttt{labrador} is classified as a specification.

The large majority of the studies find that the most frequently used modification is reformulation, followed by specification, generalization and lexical variations. \cite{Jorgensen05image,Rieh06analysis,Whittle07data,Jansen09patterns,Boldi09dango,Costa08hyponymy,Ozmutlu09markovian}. Only in \cite{He02combining} and \cite{Jones03query} almost equal numbers of reformulations and specifications are found.

The study that is most closely related to our work is the work of Huang and Efthimiadis \cite{Huang09analyzing}, who investigate the relation between modification types and clicks on search results. They found that generalization and reformulation often occur when the previous query has led to at least one click on a search result, which indicates that these modifications are mainly used after successful queries. Some types of lexical variations mainly occur when the previous query has not led to a click, suggesting a second attempt to find the same information. Specifications and reformulations appeared to be most successful: these modifications most often resulted in a click. We extend this work in three ways: 1) besides a term-based analysis, we also provide a semantic analysis of query modifications, 2) we provide a validation of our results by comparing the results of two data sets, where Huang and Efthimiadis use only one data set, 3) we explore how the results of the analysis can be used for providing feedback on users' modification strategies.

\section{Method} \label{method}
In line with previous research, we study query modifications by listing all pairs of queries that are successively entered in a search session. We will refer to the first query in a pair as the \emph{original query} and to the second query as the \emph{modified query}. Note that in sessions with more than two queries, the modified query in the one pair becomes the original query in the next pair.

We classify the relation between the queries in each pair and count the number of pairs in each class. Two classifications are used: a traditional term-based classification and a novel semantic classification.

\subsection{Term-based classification of query modifications}
For the term-based approach the queries are first stemmed using the Porter stemmer \cite{Porter80algorithm}. For each pair we determine whether, compared to the first query, in the second query terms are added (specification), removed (generalization) or replaced (reformulation). In addition, we count how many times stemming made the query identical to the previous query (lexical variation)\footnote{Consecutive queries that were identical before stemming are conflated.}. Query pairs without overlapping terms are classified as `no relation'.

\subsection{Semantic classification of query modifications}
For the semantic classification, the queries are mapped onto linked data entities (see \cite{Bizer09linked} for an overview on linked data). We make use of the \texttt{rdfs:label} property of the entities, which provides a human readable description for the entities \cite{Brickley04rdfs}. Queries are mapped on entities that have an \texttt{rdfs:label} that exactly matches the query. For instance, the query \texttt{Andre Agassi} is matched onto the entity \linebreak[4]\texttt{http://dbpedia.org/resource/Andre\_Agassi}, as this entity has the label  `Andre Agassi' (see Figure~\ref{pathSearchExampleFig}, left-hand side). 
If no exact match can be found, the queries are stemmed and mapped onto entities with labels that contain all stemmed query terms. For some queries no matching entities are found. Query pairs containing queries without matching entities are not considered in the semantic analysis.

For each pair of consecutive queries we determine the semantic relation  between the queries, as illustrated in Figure~\ref{pathSearchExampleFig}. A graph search algorithm is used for traversing links in the linked data to find the shortest series of links that connect the entities matching the two queries (their relations). 
When multiple entities match the queries, we keep only the ones for which a shortest relation is found. This often disambiguates the queries. In cases where multiple shortest relations are found, these relations are all used and the occurrence of each relation is counted as $\frac{1}{n}$, where $n$ is the number relations found.  

In the next step we abstract away from relations between specific instances and infer semantic modification types by removing the instances and keeping just the links. For instance, we may find that the relation from query \texttt{David Beckham} to query \texttt{Joe Cole} is that both refer to players in the English national football team: 

\vspace{2mm} \hspace*{10mm}\parbox{0.9\columnwidth}{\texttt{David Beckham }\textendash\texttt{DBpedia:nationalteam$\rightarrow$ \\ Eng\-land\_na\-tio\-nal\_foot\-ball\_team $\leftarrow$DBpedia:nationalteam}\textendash $\:$ \texttt{Joe Cole}}\vspace{2mm}\\ 
The arrows denote the directions of the predicates. This relation is abstracted to the modification type:

\vspace{2mm}\hspace*{10mm}\parbox{0.9\columnwidth}{\texttt{Q1 }\textendash\texttt{DBpedia:nationalteam$\rightarrow$ X $\leftarrow$DBpedia:nationalteam}\textendash \texttt{ Q2}}\vspace{2mm}

With this method for each query pair zero, one, or multiple modification types are identified.
In the last step the most likely modification types for each pair are selected by comparing the types that are found to the types of query pairs formed by randomly drawing queries from different sessions.  A complete description of the method for finding semantic modification types can be found in \cite{Hollink11semantic}.

\subsection{Quantifying modification success}
To quantify the average effectiveness of a type of query modification, we measure the success of the queries resulting from modifications of this type. Similar to \cite{Huang09analyzing}, we define a successful query as a query that is followed by at least one click on a search result. Unsuccessful queries do not result in a click and are followed directly by another query or end the search session. The motivation of this definition is that a successful query at least partially answers a user's information need. 

We define the success rate ($sr_m$) of a modification $m$ as the proportion of cases in which the modification was successful. In other words, $sr_m$ is the proportion of the query pairs with modification type $m$, where the modified query was followed by a click:
$$ sr_m = \frac{c_{m}}{c_{m}+n_{m}}$$
Here $c_m$ is the number of times $m$ was followed by a click and $n_m$ the number of times $m$ was not followed by a click. 

To compare the success rates of various modification types, we compute for each type how much using this type \emph{increases} the success rate compared to using any modification type. Formally, the increase of success rate, $isr_m$, of a modification $m$ is defined as the difference between the overall success rate and the success rate of the modification:
$$ isr_m =  \frac{\sum_{m' \in M} c_{m'}}{\sum_{m' \in M} (c_{m'}+n_{m'})} - sr_m$$
Here $M$ is the set of all modification types. 

$isr$ values are between -1 and 1. Negative values correspond to modifications with low probabilities of leading to a click (compared to other modification types), while positive values correspond to modifications with high probabilities of leading to a click. 

\section{Data sets} \label{dataSets}
We analyze query modifications in the logs of two image search engines. In image search query modification plays a larger role in than in other types of search, as users searching for images on average need more queries to find what they need \cite{Jansen04effect}. However, our analysis methods are not in any way restricted to image search.

The first data set consists of the search logs of the commercial picture portal of a European news agency. The portal provides access to more than 2 million photographic images covering a broad domain. The log files record the search interactions of professional users (mainly journalists) accessing the picture portal. We use 10 months of search logs, containing 1,094,620 queries in 520,507 sessions. Search sessions are identified using a log-in and a browser cookie and a time-out of 15 minutes. The linked data consists of various interlinked sources: the DBpedia Ontology \cite{DBpedia},  WordNet \cite{Fellbaum98wordnet,Assem06rdf}, the Cornetto Lexical Knowledge Base \cite{Vossen08integrating} (which contains both Dutch and English terms), the Getty Thesaurus of Geographical Names \cite{Getty06tgn}, and the Getty Art and Architecture Thesaurus (aat) \cite{Getty06aat}. Together these collections comprise 22 million RDF triples.

The second search engine is the search facility of the Rijksmuseum web site \cite{rijksmuseum}, a Dutch art museum. The log files cover 5.5 months and consist of 106,341 queries in 46,165 sessions, where sessions are identified using IP addresses and agent fields. As linked data, we use WordNet, Cornetto, the Dutch version of the Getty thesauri, and also various Dutch art-specific ontologies that were collected and interlinked in the E-Culture project \cite{multimedian}.

\section{Results} \label{results}
We first analyze the use of query modifications in general and determine the average success of modifications.
In Section~\ref{termbasedResults} we examine the use and successfulness of specific term-based modification types and in Section~\ref{semanticResults} of semantic modification types.

\subsection{Overall analysis} \label{topicChangesResults}

\begin{table}[b]
\caption{The overall success rate of query modifications, proportion successive query pairs for which no relation could be found (topic switches), for all query pairs (total), when the original query was successful (succ), and when the original query was unsuccessful (unsucc).}
\label{noRelationTab}
\begin{tabular*}{\columnwidth}{@{\extracolsep{\fill}}lccc} \hline
		& Total & Succ & Unsucc\\ \hline
\multicolumn{4}{c}{News photo} 		\\ \hline
success rate		& 0.34	& 0.48	& 0.27	\\
proportion no term-based relation	& 0.75	& 0.81	& 0.71	\\
proportion no semantic relation	& 0.55	& 0.58	& 0.54	\\
 \hline
\multicolumn{4}{c}{Rijksmuseum} 		\\ \hline
success rate		& 0.40	& 0.50	& 0.35	\\
proportion no term-based relation	& 0.67	& 0.77	& 0.63	\\
proportion no semantic relation	& 0.39	& 0.39	& 0.39	\\\hline
\end{tabular*} 
\end{table}

\begin{table*}[t]
\caption{Relative frequencies (freq.) and increase of success rate ($isr$) of term-based modifications, for all query pairs (total), when the original query was successful, and when the original query was unsuccessful.}
\label{termbasedResultsTab}
\begin{tabular*}{\textwidth}{@{\extracolsep{\fill}}lcccccccc} \hline
Modification		& \multicolumn{2}{c}{Total} && \multicolumn{2}{c}{After successful} && \multicolumn{2}{c}{After unsuccessful}\\ \cline{2-3} \cline{5-6} \cline{8-9}
			& freq.	& $isr$	&& freq.	& $isr$	 && freq.	& $isr$ \\ \hline
\multicolumn{9}{c}{News photo} 		\\ \hline
reformulation	& 0.48	& +0.00	&& 0.55	& $\:$--0.00	&& 0.47	& $\:$--0.00 \\
specification	& 0.32	& +0.01	&& 0.30	& $\:$--0.01	&& 0.32	& +0.02 \\
generalization	& 0.17	& $\:$--0.02	&& 0.12	& +0.04	&& 0.18	& $\:$--0.01 \\
lexical variation	& 0.03	& $\:$--0.03	&& 0.03	& $\:$--0.00	&& 0.03	& $\:$--0.03 \\ \hline
\multicolumn{9}{c}{Rijksmuseum} 		\\ \hline
reformulation	& 0.29	& +0.03	&& 0.31	& +0.05	&& 0.30	& +0.01 \\
specification	& 0.40	& $\:$--0.02	&& 0.43	& $\:$--0.05	&& 0.38	& $\:$--0.02 \\
generalization	& 0.22	& +0.01	&& 0.17	& +0.03	&& 0.24	& +0.02 \\
lexical variation	& 0.08	& $\:$--0.01	&& 0.09	& $\:$--0.02	&& 0.07	& $\:$--0.02 \\ \hline
\end{tabular*} 
\end{table*}

Table~\ref{noRelationTab} (column 2) shows the overall success rate of query modifications in the two data sets. As shown, 34\% of the News photo and 40\% of the Rijksmuseum modifications led to a click. When the original query was successful (third column), the success rate is higher than when the original query was unsuccessful (fourth column). Possibly, this difference is the result of differences between users in their ability to formulate effective queries or in their tendency to click on marginally interesting results. Another explanation could be that the successful queries stem from easier information needs. In any case, these results indicate that providing feedback on a user's search strategy is most urgent when none of the search results are clicked.

Table~\ref{noRelationTab} shows the proportion of modifications for which no  term-based type and semantic type could be found (other than `no relation'). Following previous research (e.g. \cite{Whittle07data}), we interpret the absence of a relation between queries as a shift to a new search topic. This interpretation may not always be correct as two queries for which no relation was found can still be part of the same information need and, vice versa, two queries from different information needs may accidentally be related. 
As expected, after successful queries more topic shifts occur than after unsuccessful queries. This confirms our assumption that the presence of a click can be interpreted as an indication that a user was satisfied with the result and the query was successful. Moreover, it is another indication that feedback is needed most after unsuccessful queries. 

\subsection{Term-based modifications} \label{termbasedResults}
Figure~\ref{termbasedResultFig} shows that in the News photo data the relative frequency of the four term-based modifications is in line with the majority of existing research: reformulations are used most often, followed by specifications, followed by generalizations and lexical variations \cite{Jorgensen05image,Rieh06analysis,Whittle07data,Jansen09patterns,Boldi09dango,Costa08hyponymy,Ozmutlu09markovian}. In the Rijksmuseum dataset more specifications are used than reformulations, which is also found in  \cite{He02combining,Jones03query}. Table~\ref{termbasedResultsTab} shows the relative frequencies of the modifications as well as their $isr$ scores when the orginal query was successful and when the orginal query was unsuccessful. From these data, we identify three cases where feedback on the users' modification strategies may improve search:

\begin{figure}[bht]
\centering
\includegraphics[width=0.9\columnwidth]{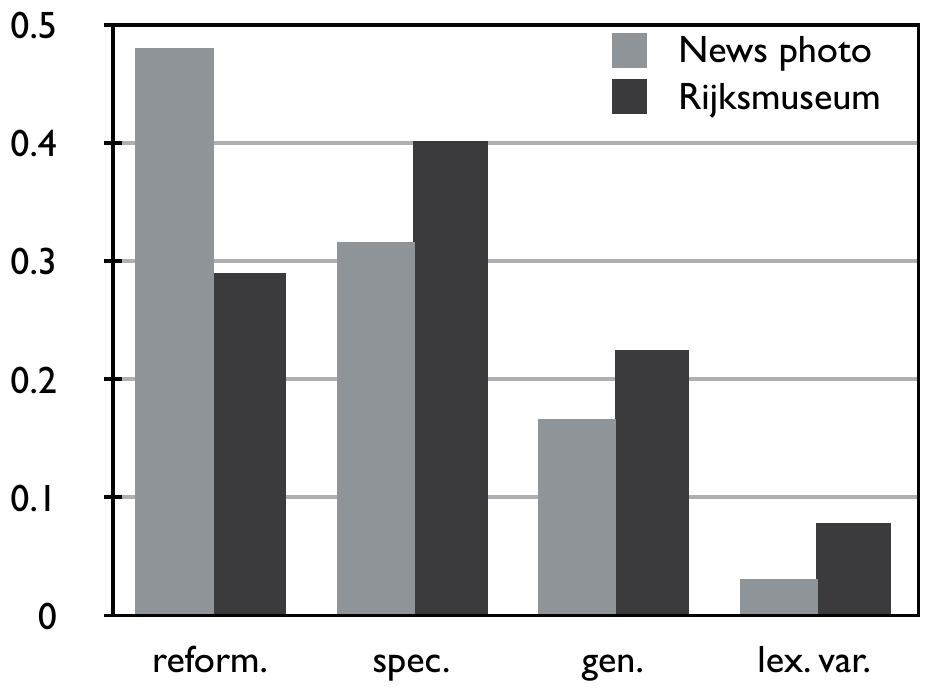}
\caption{Relative frequencies of term-based query modifications.}
\label{termbasedResultFig}
\end{figure}

\begin{description}
 \item[Feedback opportunity 1.] Generalizations are used predominantly after unsuccessful queries. Apparently, users often believe that their queries are unsuccessful because they retrieve too few interesting results (low recall) rather than because too many uninteresting results are retrieved (low precision). Even though generalizations are used mostly after unsuccessful queries, we found that they are most effective after successful queries. This reveals an opportunity for feedback: after a successful query we can advise a user that generalizing his query may lead to even more interesting results.

 \item[Feedback opportunity 2.] In contrast to generalizations, specifications are most effective after unsuccessful queries. However, on the Rijksmuseum site, where specifications play are relatively large role, they are used mainly after successful queries. Thus, after unsuccessful queries, it may be helpful to suggest a user to specify his query.

 \item[Feedback opportunity 3.] Lexical variations are less useful than average, especially after unsuccessful queries\footnote{As the search engine itself does not use stemming, lexical variations yield different search results.}. This suggests that an assistant can help users by checking whether the collection contains any not yet presented items that match queries that are after stemming identical to the users' current query. If such items are found, they can be shown directly or the corresponding lexical variations can be suggested. If no matching items are found, the assistant can inform the user  that lexical variations are not going to be effective, saving the user the effort of trying these queries.
\end{description}

\subsection{Semantic modifications} \label{semanticResults}
\begin{table*}
\centering
\caption{The ten most frequently occurring semantic modification types in the News photo data set and their increase of success rate ($isr$)}
\label{topPatternsTabNews}
\begin{tabular*}{\textwidth}{lcl}
\hline
   & $isr$	& Modification type	\\ \hline
1. & +0.02	& \texttt{\footnotesize [ ] } \\ 
2. & $\:$--0.18	&  \texttt{\footnotesize Q1 }\textendash\texttt{\footnotesize DBpedia:spouse$\rightarrow$ Q2 } \\
3. & +0.14	 	&  \texttt{\footnotesize Q1 }\textendash\texttt{\footnotesize DBpedia:nationalteam$\rightarrow$ X $\leftarrow$DBpedia:nationalteam}\textendash \texttt{ \footnotesize Q2} \\
4. & $\:$--0.07 		&  \texttt{\footnotesize Q1 }\texttt{\footnotesize $\leftarrow$DBpedia:starring}\textendash \texttt{ \footnotesize X } \textendash\texttt{\footnotesize DBpedia:starring$\rightarrow$} \texttt{ \footnotesize Q2} \\
5. & +0.08	 	&  \texttt{\footnotesize Q1 }\textendash\texttt{\footnotesize rdf:type$\rightarrow$ X $\leftarrow$rdf:type}\textendash \texttt{ \footnotesize Q2} \\
6. & $\:$--0.20	 	&  \texttt{\footnotesize Q1 }\textendash\texttt{\footnotesize DBpedia:partner$\rightarrow$ Q2 } \\
7. & $\:$--0.21	 	&  \texttt{\footnotesize Q1 $\leftarrow$aat:distinguished\_from}\textendash \texttt{ \footnotesize Q2} \\
8. & +0.04	 	&  \texttt{\footnotesize Q1 }\textendash\texttt{\footnotesize DBpedia:wordnet\_type$\rightarrow$ X $\leftarrow$DBpedia:wordnet\_type}\textendash \texttt{ \footnotesize Q2} \\
9. & +0.14	 	&   \texttt{\footnotesize Q1 }\textendash\texttt{\footnotesize DBpedia:clubs$\rightarrow$ X $\leftarrow$DBpedia:clubs}\textendash \texttt{ \footnotesize Q2} \\
10. & $\:$--0.14	 	&  \texttt{\footnotesize Q1 $\leftarrow$DBpedia:spouse}\textendash \texttt{ \footnotesize Q2} \\ \hline
\end{tabular*}
\end{table*}

\begin{table*}
\caption{Relative frequencies (freq.) and increase of success rate ($isr$) of classes of semantic modifications, for all query pairs (total), when the original query was successful, and when the original query was unsuccessful.}
\label{semanticResultsTab}
\begin{tabular*}{\textwidth}{@{\extracolsep{\fill}}lcccccccc} \hline
Modification		& \multicolumn{2}{c}{Total} && \multicolumn{2}{c}{After successful} && \multicolumn{2}{c}{After unsuccessful}\\ \cline{2-3} \cline{5-6} \cline{8-9}
			& freq.	& $isr$	&& freq.	& $isr$ && freq.	& $isr$ \\ \hline
\multicolumn{9}{c}{News photo} 		\\ \hline
sibling		& 0.19	& +0.04	&& 0.23	& +0.04	&& 0.18	& +0.01 \\
few-to-few	& 0.10	& $\:$--0.16	&& 0.07	& $\:$--0.13	&& 0.11	& $\:$--0.12 \\
same entity (\texttt{[ ]})	& 0.05	& +0.02	&& 0.04	& $\:$--0.00	&& 0.06	& +0.06 \\
other		& 0.65	& +0.01	&& 0.67	& $\:$--0.00	&& 0.65	& +0.01 \\ \hline
\multicolumn{9}{c}{Rijksmuseum} 		\\ \hline
sibling		&  0.09	& +0.01	&& 0.09	& +0.05	&& 0.08	& $\:$--0.04 \\
few-to-few	&  0.02	& $\:$--0.05	&& 0.02	& $\:$--0.02	&& 0.02	& $\:$--0.06 \\
same entity (\texttt{[ ]})	&  0.19	& $\:$--0.07	&& 0.15	& $\:$--0.10	&& 0.24	& $\:$--0.03 \\
other		&  0.70	& +0.02	&& 0.74	& +0.01	&& 0.66	& +0.02 \\ \hline
\end{tabular*} 
\end{table*}

Many different modification types emerged from the semantic query modification analysis.
The ten types that were found most frequently in the News photo data are given in Table~\ref{topPatternsTabNews}. The most common type was the \emph{same-entity} relation ([~]): two different queries that are mapped to the same entity, usually variant names for the same entity, such as \texttt{Gent} and \texttt{Gand} (the Dutch and French name of a Belgian city).  Types 2,  6, and 10 indicate that many users searched first on the name of a person and then on the name of his or her spouse or partner. Types 3, 9, and 4 respectively tell us that many users searched for related people: people who play for the same national team, who belong to the same sports club, or who star in the same movie. Types 5 and 8 both say that users searched on two entities from the same type, such as tennis players or townships. Type 7 uses the \texttt{AAT:distinguished\_from} relation from the Getty Art and Architecture Thesaurus which links closely related terms, such as \texttt{prince} and \texttt{princess}.

Inspection of the modification types revealed two important classes of modifications. \emph{Sibling relations} are modifications of the form \mbox{\texttt{Q1}\textendash\texttt{R$\rightarrow$X$\leftarrow$R}\textendash \texttt{Q2}} or \mbox{\texttt{Q1}\texttt{$\leftarrow$R}\textendash \texttt{X}\textendash\texttt{R$\rightarrow$}\texttt{Q2}}. Examples include types 3, 4, 5, 8, and 9 in Table~\ref{topPatternsTabNews}. This shows that many people searched for two entities with some common property, such as two actors starring in the same movie or two hyponyms of a WordNet concept. The second frequently occurring class of modifications are \emph{direct few-to-few relations}, which are defined as a relaxed version of one-to-one relations, where `few' means on average less than 2. Examples of direct few-to-few relations are `spouse' and `has-capital'. 

Table~\ref{semanticResultsTab} shows the relative frequency and $isr$ of the classes of semantic modifications. The $isr$ scores of the most frequent individual modification types are given in Table~\ref{topPatternsTabNews}. We identify three more opportunities for feedback:

\begin{description}
 \item[Feedback opportunity 4.]  Sibling modifications are on average relatively successful (see Table~\ref{semanticResultsTab} and types 3, 5, and 9 in Table~\ref{topPatternsTabNews}). Some siblings, however, are not very successful, such as type 4. Few-to-few relations are often unsuccessful (e.g. types 2, 6 and 10). The success of these individual modification types can be used for giving detailed feedback. An assistant could, for instance, advise a user searching for a soccer player to try the name of another player from the same team.

 \item[Feedback opportunity 5.]  Sibling relations are most effective after successful queries. This findings can be used to make recommendations dependent on the success of the previous query, i.e. to recommend sibling relations only when a click is made.  Note, however, that many users are already able to use sibling modifications strategically, as these modifications are found predominantly after successful queries.

 \item[Feedback opportunity 6.]  Like lexical variations, same-entity modifications are not very effective after successful queries. Same-entity modifications can be used to provide feedback in the same manner as lexical variations by checking whether the collection contains items that match the same entity as the user has searched for, but that are described by different terms.
\end{description}



\section{Conclusions}
Our analyses revealed that users are not always able to make effective query modifications.
We identified six situations in which feedback on a user's query modification strategy may improve the search. Using our findings a feedback assistant can advise a user in some situation to use or not to use particular term-based or semantic query modifications.

Comparing the semantic and the term-based analyses, we found that the insights that they provide are complementary. Both approaches yielded unique opportunities for feedback. As far as the findings of the two analysis overlap, their results are consistent.

In this paper we studied the relation between clicks on search results and the ways users modify their queries. So far, we did not take into account \emph{why} a set of search results was (un)satisfactory for a user. Did it contain too few results? Did it contain results not related to the user's information need? In the next step of our research we will explore the influence of the ambiguity, specificity, and size of the result set on query modifications.


\bibliographystyle{abbrv}
\bibliography{USEWOD}

\end{document}